\documentclass{aa}
\usepackage{graphicx}

\def\jh{\mbox{$\rm (J-H)$}}
\def\mM{\mbox{$\rm (m-M)_0$}}
\def\ebv{\mbox{$\rm E(B-V)$}}
\def\ejh{\mbox{$\rm E(J-H)$}}

\def\rc{\mbox{$\rm R_{core}$}}
\def\rt{\mbox{$\rm R_{tidal}$}}
\def\rlim{\mbox{$\rm R_{lim}$}}
\def\ms{\mbox{$\rm M_\odot$}}
\def\ds{\mbox{$\rm d_\odot$}}
\def\dgc{\mbox{$\rm d_{GC}$}}
\def\mj{\mbox{$\rm M_J$}}
\def\jj{\mbox{$\rm J$}}
\def\hh{\mbox{$\rm H$}}
\def\mMs{\mbox{$\rm m_{MS}$}}
\def\mobs{\mbox{$\rm m_{obs}$}}
\def\mtot{\mbox{$\rm m_{tot}$}}

\def\age{\mbox{$\rm\tau_{age}$}}
\def\feH{\mbox{$\rm [Fe/H]$}}
\def\fb{\mbox{$\rm f_{bin}$}}
\def\cmf{\mbox{$\rm \chi_{imf}$}}
\def\tr{\mbox{$\rm t_{relax}$}}
\def\tc{\mbox{$\rm t_{cross}$}}

\begin{document}

\title{Spatial dependence of 2MASS luminosity and mass functions in the old 
open cluster NGC\,188}

\author{C. Bonatto\inst{1}, E. Bica\inst{1} \and J.F.C. Santos Jr\inst{2}}

\offprints{Ch. Bonatto - charles@if.ufrgs.br}

\institute{Universidade Federal do Rio Grande do Sul, Instituto de F\'\i sica, 
CP\,15051, Porto Alegre 91501-970, RS, Brazil\\
\mail{}
\and
Universidade Federal de Minas Gerais, ICEx, Departamento de F\'\i sica,
CP\,702, Belo Horizonte 30162-970, MG, Brazil
}

\date{Received --; accepted --}

\abstract{Luminosity and mass functions in the old open cluster NGC\,188 are analysed 
by means of J and H 2MASS photometry, which provides uniformity and spatial coverage 
for a proper background subtraction. With an age of about 6--8\,Gyr, NGC\,188 is expected 
to be suffering the effects of advanced dynamical evolution. Indeed, previous works in 
optical bands have suggested the presence of mass segregation. Within the uncertainties,
the observed projected radial density profile of NGC\,188 departs from the two-parameter 
King model in two inner regions, which reflects the non-virialized dynamical state and 
possibly, some degree of non-sphericity in the spatial shape of this old open cluster.
Fits with two and three-parameter King models to the radial distribution of stars resulted 
in a core radius $\rc=1.3\pm0.1$\,pc and a tidal radius $\rt=21\pm4$\,pc, about twice
as large as the visual limiting radius. The concentration parameter $c=1.2\pm0.1$ of 
NGC\,188 makes this open cluster structurally comparable to the loose globular clusters. 
The present 2MASS analysis resulted in significant slope variations with distance in 
the mass function $\phi(m)\propto m^{-(1+\chi)}$, being flat in the central parts 
($\chi=0.6\pm0.7$) and steep in the cluster outskirts ($\chi=7.2\pm0.6$). The overall 
mass function has a slope $\chi=1.9\pm0.7$, slightly steeper than a standard Salpeter
mass function. In this context, NGC\,188 is similar to the 3.2\,Gyr,  dynamically evolved
open cluster M\,67. Solar metallicity Padova isochrone fits to the near-infrared colour-magnitude 
diagram of NGC\,188 resulted in an age of $7.0\pm1.0$\,Gyr. The best fit, obtained with the 
7.1\,Gyr isochrone, produced a distance modulus $\mM=11.1\pm0.1$, $\ebv=0.0$, and a distance 
to the Sun $\ds=1.66\pm0.08$\,kpc. The observed stellar mass (in the range $0.98\,\ms - 
1.08\,\ms$) in NGC\,188 is $\mobs=380\pm12\,\ms$. A simple extrapolation of the observed 
overall mass function to stars with $0.08\,\ms$ resulted in a total present mass of 
$\mtot\sim(1.8\pm0.7)\times10^4\,\ms$. On the other hand, for a more realistic initial 
mass function which flattens in the low-mass range, the total mass in NGC\,188 drops to 
$\mtot\sim(3.8\pm1.6)\times10^3\,\ms$.  Since mass-loss processes such as evaporation 
and tidal stripping have been occurring in this old open cluster for about 7\,Gyr, the 
primordial mass in NGC\,188 must have been significantly larger than $\sim4\times10^3\,\ms$.
We also examined the consequences of the presence of unresolved binaries and concluded that,  
even if dominant in the CMD, binaries alone are not responsible for the flat central mass 
function, which supports the mass-segregation scenario.

 }

\titlerunning{Dynamical state of NGC\,188}

\authorrunning{C. Bonatto et al.}

\maketitle

\section{Introduction}
\label{intro}

With an age of about 7\,Gyr (e.g. Carraro \& Chiosi \cite {CarChi1994}), NGC\,188 is one of 
the oldest known open clusters in the Galaxy. Located relatively far from the disk, its field 
is neither heavily contaminated by background stars nor obscured by dust, and contains a few 
hundreds of member stars. These conditions make NGC\,188 an excellent laboratory for 
testing modern theories of stellar and dynamical evolution of star clusters. 

The main  effects related to the dynamical evolution of an open cluster are {\it (i)}
mass segregation, {\it (ii)} tidal stripping and disruption by the Galactic gravitational 
field, and {\it (iii)} interactions with the disk, since a typical open cluster at the 
solar radius will cross the Galactic plane 10--20 times before being disrupted (de la 
Fuente Marcos \cite{delaF1998}).  Evaporation of a cluster may result from two-body 
relaxation processes, in which several stars can acquire positive energy during the 
closest approaches and eventually leave the cluster. Mass-loss during the 
course of stellar evolution and encounters with giant molecular clouds (e.g. Wielen 
\cite{Wielen1991}) also contribute to reducing a cluster's lifetime. Considering all 
these factors, the destruction time-scale for open clusters in the solar neighbourhood 
is about 600\,Myr (Bergond, Leon, \& Guibert \cite{Bergond2001}). Consequently, only the 
most massive open clusters or those located at large Galactic radii are expected to 
survive longer than a few Gyr (Friel \cite{Friel1995}).  Indeed, de la Fuente Marcos 
(\cite{delaF1998}) has shown that the Milky Way should host several hundred thousands of 
open cluster remnants which have been disrupted both by the Galaxy and their own dynamical 
evolution.

Taking into account all of  the above factors, as well as its location in the 
Galaxy, the dynamical evolution of NGC\,188 should  be essentially associated to mass 
segregation,  evaporation and tidal stripping by the Galactic gravitational field.

As a consequence of mass segregation, the central regions of the more evolved clusters
should present a main sequence (MS) depleted of low-mass stars, thus creating a core 
rich in compact and giant stars (Takahashi \& Portegies Zwart \cite{TakaP2000}). Also, 
the luminosity (or mass) function (hereafter LF and MF) of a dynamically evolved 
open cluster should present slope changes with distance to the center, being flatter 
in the central parts than in the outer regions. 

Examples of advanced dynamical evolution in star clusters have been reported in recent 
years, e.g. in NGC\,3680 (Anthony-Twarog et al. \cite{Twa1991}; Bonatto, Bica \& Pavani 
\cite{BBP2004}) and M\,67 (Bonatto \& Bica \cite{BB2003}), in which the presence of a 
core depleted of low-MS stars and a  halo rich in low-mass stars have been confirmed 
with 2MASS photometry. 

The accurate determination of LF spatial variations, as well as the characterization
of a cluster's stellar content, depends critically on a proper background subtraction. 
The uniform and essentially complete sky coverage provided by 2MASS (Two Micron All Sky 
Survey, Skrutskie et al. \cite{2mass1997}) has proven to be a powerful tool for properly 
taking into account background regions with suitable star count statistics (see, e.g. Bonatto 
Bica \& Pavani \cite{BBP2004}; Bonatto \& Bica \cite{BB2003}; Bica, Bonatto \& Dutra 
\cite{BBD2003}; Bica, Bonatto \& Dutra \cite{BBD2004}).

In the present work we investigate the dynamical state and stellar content of NGC\,188.
We employ J and H 2MASS photometry to analyse a large spatial area in the direction of 
the cluster. In particular, we search for spatial changes in the LF which would 
characterize the advanced dynamical state of this open cluster. This is the first analysis 
of NGC\,188 in the near-infrared with wide spatial coverage and resolution.

This paper is organized as follows. In Sect.~\ref{N188} we provide general data on NGC\,188
and present a brief review on this open cluster. In Sect.~\ref{2massPh} we obtain the 2MASS 
photometry and introduce the $\jj\times\jh$ colour-magnitude diagrams (CMDs). In Sect.~\ref{StructAnal} we discuss 
the projected radial density distribution of stars and derive structural parameters for NGC\,188. 
In Sect.~\ref{Fund_par} we fit isochrones to the near-infrared CMD and derive cluster 
parameters. In Sect.~\ref{LMF} we derive the LFs and MFs, estimate the stellar mass still 
present in NGC\,188 and examine the presence of unresolved binaries in the CMD. Concluding 
remarks are given in Sect.~\ref{Conclu}. In Appendix~\ref{cmf} we discuss the consequences of 
applying colour-filters to the CMD and the background subtraction. 

\section{The old open cluster NGC\,188}
\label{N188}

NGC\,188 has been in the spotlight since the early works of Sandage (\cite{Sandage1961},
\cite{Sandage1962}) when the age of this open cluster was estimated to be $\age\sim10-16$\,Gyr, 
making NGC\,188 the probably oldest observable open cluster in the Galaxy. However, 
later works revised its age downwards (see Carraro et al. \cite{Carraro1994} for a comprehensive 
age census) to the presently accepted value of around 6--7\,Gyr (von Hippel \& Sarajedini 
\cite{Hippel1998}, Sarajedini et al. \cite{Sarajedini1999}). The WEBDA open cluster 
database\footnote{\em http://obswww.unige.ch/webda} (Mermilliod \cite{Merm1996}) gives 
$\age\approx4.3$\,Gyr for NGC188. 

The actual, advanced age of NGC\,188 still makes it one of the oldest surviving open clusters 
in the Milky Way. As such, NGC\,188 has been used to test stellar evolution models, since 
its CMD exhibits a wide giant branch (e.g. Twarog \cite{Twarog1978}, Norris \& Smith 
\cite{NorSm1985}), blue stragglers (e.g. Leonard \& Linnell \cite{LeoLin1992}, Dinescu 
et al. \cite{Dinescu1996}) and numerous binaries (von Hippel \& Sarajedini 
\cite{Hippel1998}, and references therein). Most of the previous research has been restricted 
to optical bands.

One of the reasons why NGC\,188 has survived to such an advanced age may be its almost 
circular, highly inclined external ($\rm 9.5\leq R(kpc)\leq11$) orbit, which avoids the inner 
disk regions for most of the time ( a plot of the Galactic orbit of NGC\,188 is shown 
in Fig.~1 of Carraro \& Chiosi \cite {CarChi1994}). As a consequence of its orbit, the 
probability of significant dynamical interactions with giant molecular clouds becomes 
considerably reduced and thus, the r\^ole of internal processes in the dynamical evolution 
of NGC\,188 increases in importance.

Based on V and I-band photometry of the central parts of NGC\,188, (von Hippel \& Sarajedini 
\cite{Hippel1998}), and UBVRI CCD photometry, Sarajedini et al. (\cite{Sarajedini1999}), 
suggested that NGC\,188 has a typical initial mass function. But, as a consequence of the 
advanced, i.e. mass-segregated dynamical state, the low-mass stars presently missing from
the central parts are either in the cluster outskirts or have left the cluster 
(von Hippel \& Sarajedini \cite{Hippel1998}).

WEBDA gives a slightly subsolar metallicity for NGC\,188, which agrees with other 
values, e.g. $\feH=0.02\pm0.11$ (Caputo et al. \cite{Caputo1990}), $\feH=-0.12\pm0.16$ (Hobbs, 
Thorburn \& Rodriguez-Bell \cite{Hobbs1990}) and $\feH=-0.02$ (Twarog, Ashman \& Anthony-Twarog
\cite{Twa1997}).

The central coordinates of NGC\,188, precessed to J2000, are $\alpha=00^h\,47^m\,53^s$\ and 
$\delta=+85^\circ\,15\arcmin\,30\arcsec$, which convert to $\ell=122.85^\circ$\ and 
$b=+22.39^\circ$.

\section{The 2MASS photometry}
\label{2massPh}

Despite having lost its title as the oldest Galactic open cluster, NGC\,188 still deserves 
attention, since its age implies advanced stellar and dynamical evolution, and its 
location in the Galaxy favours deep observations, particularly in the infrared. We 
base the present analysis of NGC\,188 essentially on J and H 2MASS\footnote{All Sky data 
release, available at {\em http://www.ipac.caltech.edu/2mass/releases/allsky/}} photometry. 
2MASS photometric errors typically attain 0.10\,mag at $\jj\approx16.2$ and $\hh\approx15.0$, 
see e.g. Soares \& Bica (\cite{SB2002}). The 
VizieR\footnote{\em http://vizier.u-strasbg.fr/viz-bin/VizieR?-source=II/246} tool has been 
used to extract stars in a circular area of 45\arcmin\ radius centered on the coordinates 
given in Sect.~\ref{N188}.
In order to maximize the statistical significance and representativeness of background star 
counts, we used the stars in the outermost annulus ($\rm 40\arcmin\leq R\leq45\arcmin$) as offset field. 
See Appendix~\ref{cmf} for a discussion on background subtraction.

\begin{figure}
\resizebox{\hsize}{!}{\includegraphics{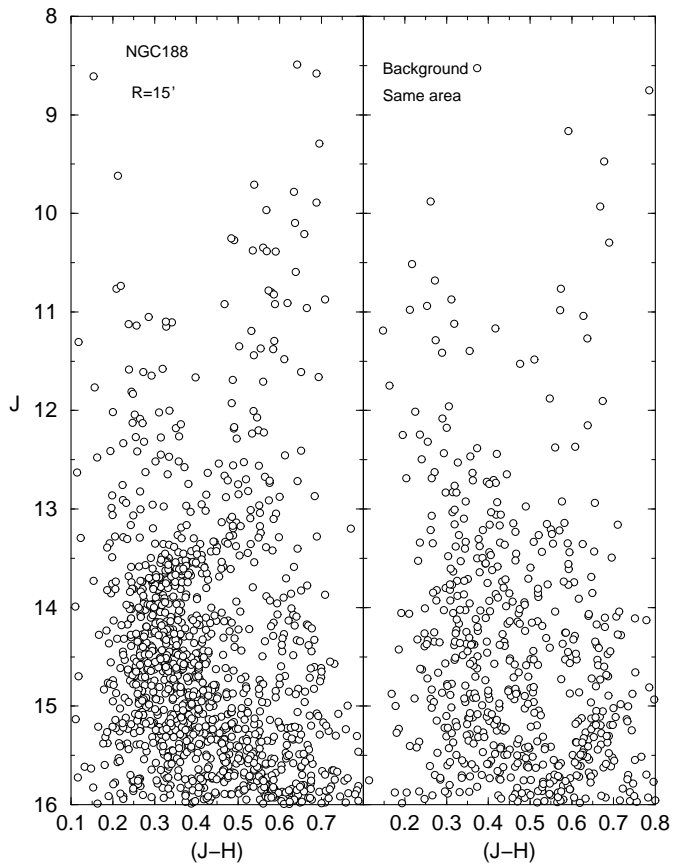}}
\caption[]{$\jj\times\jh$ CMDs for the cluster NGC\,188 (left panel, containing 1804
stars) and corresponding offset field (right panel,  containing 998 stars). Both CMDs 
have been extracted within equal areas. }
\label{fig1}
\end{figure}

In Fig.~\ref{fig1} we show the $\jj\times\jh$ CMDs for NGC\,188 (left panel) along with 
the corresponding (equal area) offset field (right panel). In order to maximize visualization 
of star sequences, the CMD in the direction of NGC\,188 has been built with stars extracted 
within an area of 15\arcmin\ in radius. This dimension is, however, smaller than the limiting 
radius of NGC\,188 (Sect.~\ref{StructAnal}). 

The near-infrared CMD of NGC\,188 presents a well-defined MS with several stars occupying the 
turnoff, as well as a wide giant branch. The large width of the MS, which cannot be entirely
accounted for by photometric errors (Sect.~\ref{photom}), indicates that this cluster
may contain a large fraction of binaries, confirming previous optical observations 
(von Hippel \& Sarajedini \cite{Hippel1998}, and references therein). These features are not 
present in the CMD of the comparison field. The nearly vertical sequence at $0.2\leq\jh\leq0.4$
and $10.5\leq\jj\leq13.5$, 
present in both CMDs but more populated in NGC\,188, is probably due to disk stars and, to some 
extent, to the presence of blue stragglers in NGC\,188 (Leonard \& Linnell \cite{LeoLin1992}, 
Dinescu et al. \cite{Dinescu1996}).

The number of stars in the CMD in the direction of NGC\,188 
is 1804, while in its comparison field this number drops to 998. 

\section{Cluster structure}
\label{StructAnal}

\begin{figure} 
\resizebox{\hsize}{!}{\includegraphics{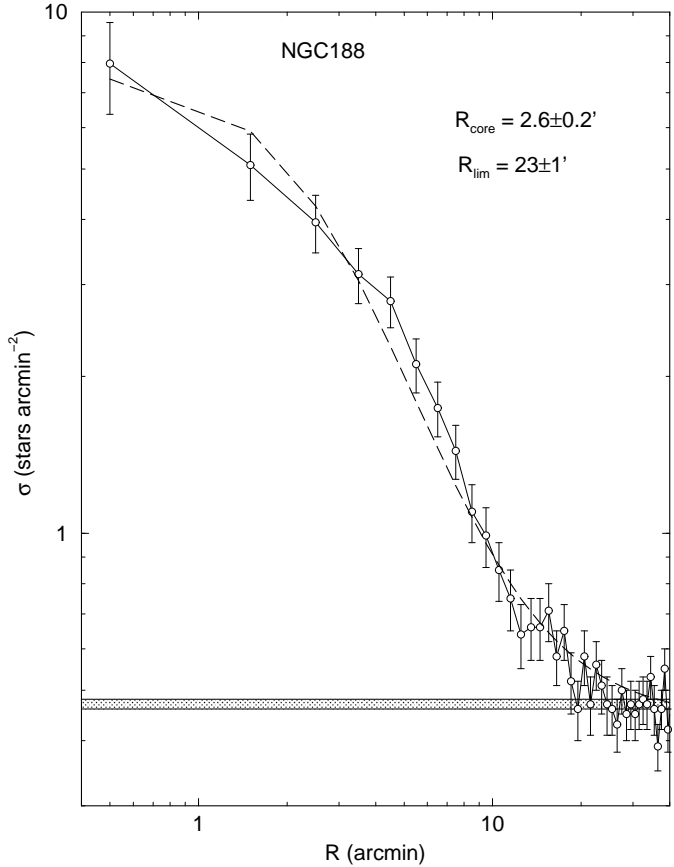}}
\caption[]{Projected radial distribution of surface star density. The average background level
is shown as a narrow shaded area; $1\sigma$ Poisson errors are also shown. Magnitude 
($\jj<16.0$) and colour cutoffs have been applied to the object and offset field. The dashed line 
shows a surface density profile (two-parameter King model) fit to the radial distribution of stars.}
\label{fig2}
\end{figure}

The overall cluster structure is analysed by means of the star density radial distribution, 
defined as the projected number of stars per area in the direction of a cluster,
shown in Fig.~\ref{fig2} for NGC\,188. 

Before counting stars, we applied a cutoff ($\jj<16.0$) to both cluster and offset field 
to avoid oversampling, i.e. spatial variations in the number of faint stars and spurious 
detections. Colour filters (Appendix~\ref{cmf}) have also been applied to both cluster and 
corresponding offset field, in order to account for the contamination by the Galaxy. This 
procedure has been previously applied in the analysis of the open cluster M\,67 (Bonatto \& Bica 
\cite{BB2003}). As a result of the filtering process, the number of stars in the direction 
of NGC\,188 inside an area of 20\arcmin\ in radius turns out to be 1201, compared to 
583 in its offset field (same area). The radial distribution has been determined by counting 
stars inside concentric annuli with a step of 1.0\arcmin\ in radius. The background contribution 
level corresponds to the average number of stars included in the external annulus, which lies
outside the cluster border at $\rm 40\arcmin\leq R\leq 45\arcmin$.

The projected radial density profile of NGC\,188 is very smooth with a well-defined central 
concentration of stars. The relatively large number of stars in this cluster produces small 
$1\sigma$ Poisson error bars along most of the profile. Considering the profile fluctuations 
with respect to the background level, we can define for NGC\,188 a limiting radius of 
$\rlim=24\pm1\arcmin$. Because of the essentially null-contrast between cluster and background star density, 
any statistical analysis for regions beyond $\rlim$ would produce excessively high Poisson error 
bars, and, consequently, meaningless results. Thus, for practical purposes, we can consider that most 
of the cluster's stars are contained within $\rlim$.

First order structural parameters of NGC\,188 are derived by fitting the two-parameter King 
(\cite{King1966}) surface density profile to the background-subtracted radial distribution of 
stars. The two-parameter King model essentially describes the central region 
of normal clusters (King \cite{King1966b}; Trager, King \& Djorgovsky \cite{TKD95}). The fit 
has been performed using a non-linear least-squares fit 
routine which uses the error bars as weights. The best-fit solution is shown in Fig.~\ref{fig2} 
as a dashed line, and the resulting core radius is $\rc=2.6\pm0.2\arcmin$. With a distance to 
the Sun $\ds=1.66\pm0.08$\,kpc (Sect.~\ref{Fund_par}), the linear core radius of NGC\,188 turns 
out to be $\rc=1.3\pm0.1$\,pc and the angular diameter ($2\times\rlim$) of $48\pm2\arcmin$ 
converts to a linear limiting diameter of $23.2\pm1.5$\,pc. 

Within the errors, the observed density profile departs from the two-parameter King model 
in opposite senses, particularly for the regions $\rm 0.9\leq R(\arcmin)\leq3.5$ and $\rm 3.5\leq 
R(\arcmin)\leq8.6$ (Fig.~\ref{fig2}). Considering the physical premises of the King model, 
these deviations may be a consequence of the non-virialized dynamical state and possibly, 
some degree of non-sphericity in the spatial shape of NGC\,188.

The core radius of NGC\,188 is very similar to that of the $\age=3.2$\,Gyr,
relaxed open cluster M\,67 ($\rc=1.2\pm0.1$\,pc -- Bonatto \& Bica 
\cite{BB2003}).  However, its limiting diameter is larger than that of
M\,67 ($12.7\pm0.6$\,pc), probably a consequence of NGC\,188's older age, more 
advanced dynamical state and larger mass (Sect.~\ref{MassF}).

\subsection{Tidal radius}
\label{tidal_R}

The rather massive (Sect.~\ref{MassF}) nature and present large galactocentric distance 
(Sect.~\ref{Fund_par}) of NGC\,188 make it possible to fit its projected radial distribution 
of stars with the more complete empirical density law of King (\cite{King1962}), which includes 
the tidal radius $\rt$ as parameter in the model. The tidal radius depends both on the effect 
of the Galactic tidal field on the cluster and the subsequent internal relaxation and dynamical
evolution of the cluster (Allen \& Martos \cite{AlMart88}). In particular, Oh \& Lin 
(\cite{OH1992}) have shown, through numerical simulations, that relaxation affects the tidal 
radius formation. In the case of NGC\,188, the tidal radius determination is made possible 
by the spatial coverage and uniformity of 2MASS photometry, which allows one to obtain reliable 
data on the projected distribution of stars for large extensions around clusters. 

The three-parameter model of King (\cite{King1962}) is a potential source of information 
both on the internal structure of the cluster, by providing reliable determinations of the 
core radius and the central density of stars, and on the Galactic tidal field, since the tidal 
radius is linked to the stripping of stars from the cluster by the Galactic tidal field. Since 
the three-parameter King model provides a better description of the outer regions of a cluster
than the two-parameter model, it is particularly suitable for deriving a reliable value of 
$\rt$.

\begin{figure} 
\resizebox{\hsize}{!}{\includegraphics{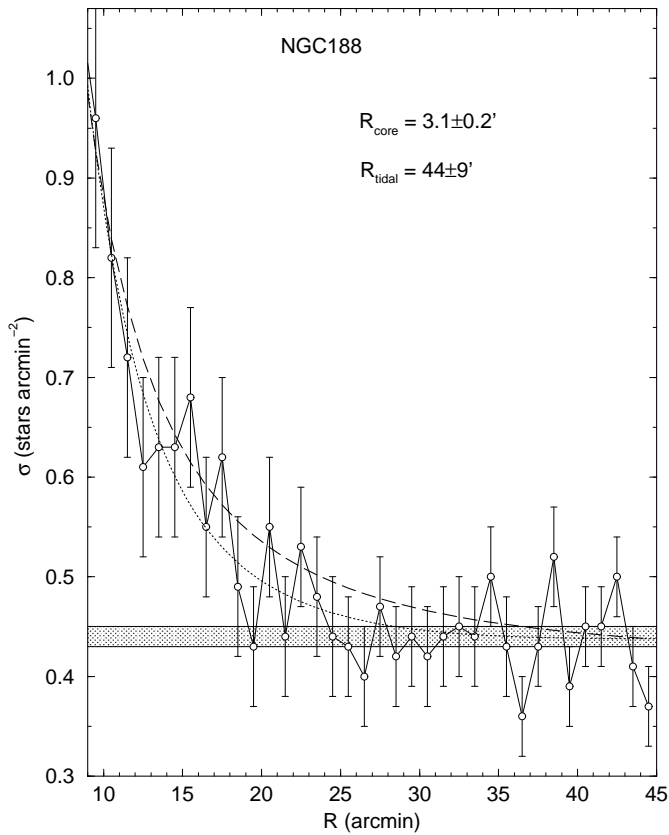}}
\caption[]{Fit of King models to the projected radial distribution of stars of NGC\,188. Dashed 
line: two-parameter model; dotted line: three-parameter model. }
\label{fig3}
\end{figure}

The best-fit solution, with a correlation coefficient of 0.96 and an RMS error of 0.83, 
resulted in $\rc=3.1\pm0.2\arcmin$ and $\rt=44\pm9\arcmin$, which 
convert to $\rc=1.5\pm0.1$\,pc and $\rt=21\pm4$\,pc. This solution is shown in
Fig.~\ref{fig3} for regions more external than $\rm R=8\arcmin$. The two-parameter model
fit is also shown in Fig.~\ref{fig3} for comparison purposes.

The present value of $\rt$ is nearly twice as large as that estimated by Keenan, Innanen \& House 
(\cite{KIH73}) and, within the errors, comparable to the theoretical value calculated by Allen \& 
Martos (\cite{AlMart88}). Interestingly, we note that the physical structure of NGC\,188 extends 
well beyond the visual limits, since the presently derived value of $\rt$ is about twice that of 
$\rlim$. We note as well that the inclusion of the additional parameter $\rt$ in the fit produced 
an increase of about 15\% in $\rc$, a value which differs $\approx2\sigma$ from that derived 
with the two-parameter King model. However, since the two-parameter King model appears to give a
better description of the inner regions of star clusters (King \cite{King1966b}; Trager, King \& 
Djorgovsky \cite{TKD95}), we will adopt the previous value of $\rc=1.3\pm0.1$\,pc as the core 
radius of NGC\,188. With the values derived from the three-parameter model, the concentration 
parameter of NGC\,188 turns out to be $c=\log\left(\frac{\rt}{\rc}\right)=1.2\pm0.1$. To check 
how NGC\,188 fits in the context of evolved star clusters, we compare its concentration parameter 
with those of 141 Galactic globular clusters (compiled by Harris \cite{H96})\footnote{\em\small 
http://physwww.physics.mcmaster.ca/\%7Eharris/mwgc.dat}, which also includes the globular 
cluster sample of Trager, King \& Djorgovsky (\cite{TKD95}). We conclude that NGC\,188 is 
structurally comparable to the loose globular clusters, but still more concentrated than 
$\approx25\%$ of the sample. 

\section{Fundamental parameters}
\label{Fund_par}

To maximize cluster membership probability, the analyses in the following two sections 
will be restricted to stars extracted within \rlim\ (Sect.~\ref{StructAnal}). Cluster 
parameters will be derived using solar metallicity Padova isochrones from Girardi et al. 
(\cite{Girardi2002}) computed with the 2MASS J, H, and K$_S$ filters\footnote{Available at 
{\em http://pleiadi.pd.astro.it}}. The 2MASS transmission filters produced isochrones 
very similar to the Johnson ones, with differences of at most 0.01 in \jh\ (Bonatto, Bica 
\& Girardi \cite{BBG2004}). The solar metallicity isochrones have been selected to be 
consistent with previous results (e.g. Caputo et al. \cite{Caputo1990}, Hobbs, Thorburn 
\& Rodriguez-Bell \cite{Hobbs1990}, and Twarog, Ashman \& Anthony-Twarog \cite{Twa1997}). 
For reddening and absorption transformations we use $\rm R_V=3.2$, and the relations 
$\rm A_J=0.276\times A_V$ and $\ejh=0.33\times\ebv$, according to Dutra, Santiago \& Bica 
(\cite{DSB2002}) and references therein. 

\begin{figure}
\resizebox{\hsize}{!}{\includegraphics{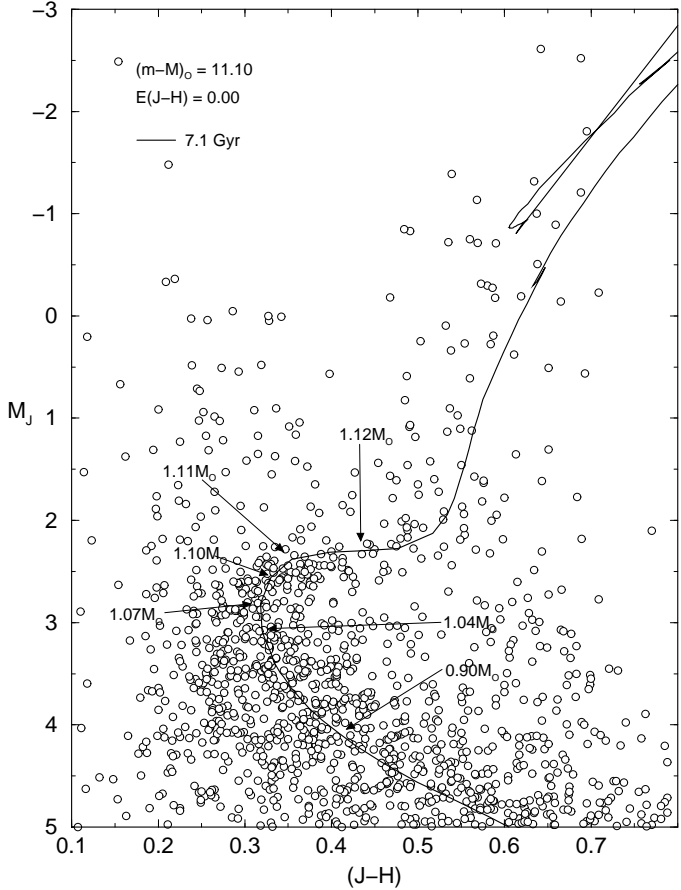}}
\caption[]{Isochrone fit to the $\mj\times\jh$ CMD of NGC\,188 with the $\age=7.1$\,Gyr
Padova isochrone, resulting in $\mM=11.1\pm0.1$, $\ebv=0.0$ and $\ds=1.66\pm0.08$\,kpc. 
Representative stellar masses are indicated.}
\label{fig4}
\end{figure}

The presence of a well-defined turnoff and giant branch in the $\mj\times\jh$ CMD of 
NGC\,188 constrain the age solution to the 6.3\,Gyr, 7.1\,Gyr and 8.0\,Gyr Padova 
isochrones. The best fit, shown in Fig.~\ref{fig4}, corresponds to the $\age=7.1$\,Gyr 
solution. Accordingly, we adopt as the age of NGC\,188 $\age=7.0\pm1.0$\,Gyr. The \mj\ 
values are obtained after applying the distance modulus 
derived below for NGC\,188. From the isochrone fit and related uncertainties we derive
a distance modulus $\mM=11.1\pm0.1$, $\ebv=0.0$ and a distance to the Sun 
$\ds=1.66\pm0.08$\,kpc. For illustrative purposes, we indicate representative stellar 
masses along the isochrone.

With the above distance to the Sun, the Galactocentric distance of NGC\,188 is
$\dgc=8.9\pm0.1$\,kpc.

At this point it may be useful to estimate the expected mass of NGC\,188 based on
its  Galactocentric distance and tidal radius. For clusters with a nearly circular
orbit we can use the relation $\rm M_c\approx3M_G\left(\frac{\rt}{R_p}\right)^3$ (King
\cite{King1962}), where $\rm R_p$ is the perigalacticon distance and $\rm M_G$ is the Galaxy 
mass inside $\rm R_p$. For the nearly circular orbit of NGC\,188, $\rm R_p\approx d_{GC}$, and 
at this distance, $\rm M_G\approx1\times10^{11}\,\ms$ (Carraro \& Chiosi \cite{CarChi1994}). 
With these values, the expected mass of NGC\,188 turns out to be 
$\rm M_c\approx(3.9\pm2.2)\times10^3\,\ms$. 

\section{Luminosity and mass functions}
\label{LMF}

In this section we analyze the spatial dependence of the luminosity and mass functions 
in NGC\,188, which will be subsequently used to derive the stellar mass still locked in 
this old open cluster. 

\subsection{Luminosity function}
\label{LumF}

The 2MASS uniform sky coverage has been shown to be an ideal tool for studies like the 
present one, since the entire cluster area can be included in the analyses, and representative 
offset fields can be selected around the cluster (e.g. Bonatto \& Bica \cite{BB2003}, Bica, 
Bonatto \& Dutra \cite{BBD2003}). As a consequence, advanced stages of mass segregation and 
the cumulative effects of the Galactic tidal pull can be properly detected and interpreted.

\begin{figure}
\resizebox{\hsize}{!}{\includegraphics{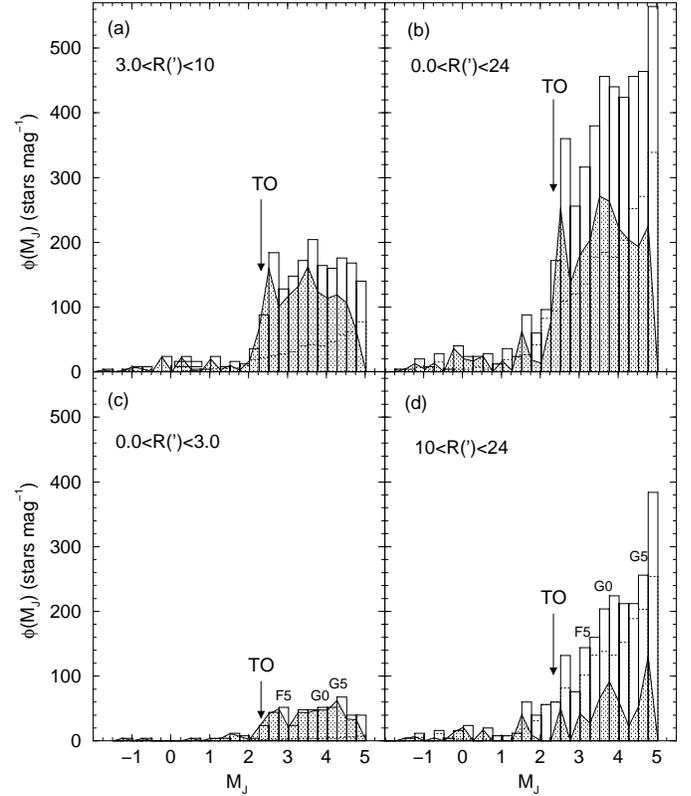}}
\caption[]{Luminosity functions ($\phi(\mj)$) in terms of the absolute magnitude \mj. 
Magnitude ($\jj<16.0$) and colour cutoffs have been applied to the object and offset 
fields. Continuous line: star counts in the cluster area; dotted line: star counts 
in the offset field; shaded area: background-subtracted LF. The turnoff (TO) is 
indicated in each panel.}
\label{fig5}
\end{figure}

In Fig.~\ref{fig5} we show the LFs ($\phi(\mj)$) in the J filter (shaded area), built 
as the difference in the number of stars in a given magnitude bin between object 
(continuous line) and offset field (dotted line). The LFs are given in terms of the 
absolute magnitude \mj, obtained after applying the distance modulus derived in 
Sect.~\ref{Fund_par}. The bin in magnitude is $\Delta\mj=0.25$\,mag. The LFs in 
Fig.~\ref{fig5} are built after applying magnitude ($\jj<16.0$) and colour cutoffs to 
the object and offset field (Sect.~\ref{Fund_par}). 

The large projected area of NGC\,188, containing $\sim620$ member stars 
(Sect.~\ref{StructAnal}), can be used to search for spatial variations in the stellar 
content, with statistically significant results. Thus, LFs for different regions inside 
NGC\,188 have been built following the structures present in the radial density profile 
(Fig.~\ref{fig2}). These LFs are shown in panels (a) to (d) of Fig.~\ref{fig5}. In 
particular, the LF in panel (c) corresponds to the core region, while in panel (b) we 
show the overall ($\rm 0.0\arcmin\le R\le\rlim$) LF. The background $\phi(\mj)$ has been 
scaled to match the projected area of each region. Representative MS spectral types 
(adapted from Binney \& Merrifield \cite{Binney1998}) are shown in panels (c) and (d). 
The turnoff ($\mj\approx2.27$, spectral type $\approx$F\,3, mass $\approx1.10\,\ms$) is 
indicated in all panels. The excess at $\mj\approx2.5$ corresponds to the accumulation 
of stars near the turnoff (Fig.~\ref{fig4}). Hereafter we will refer to the maximum
number count in the LFs as turnover.

As expected from the long-term dynamical evolution of this old open cluster, the large 
number of member stars contained inside R\,$=24\arcmin$ (panels (a), (c) and (d)), does 
not present a uniform spatial distribution of MS spectral types (magnitude range 
$3.0\leq\mj\leq4.5$). The region with $\rm R\geq10\arcmin$ seems to have an LF steeper than 
those of the more internal ones, the core region in particular, which is characteristic
of mass segregation.

\subsection{Mass function}
\label{MassF}

\begin{figure}
\resizebox{\hsize}{!}{\includegraphics{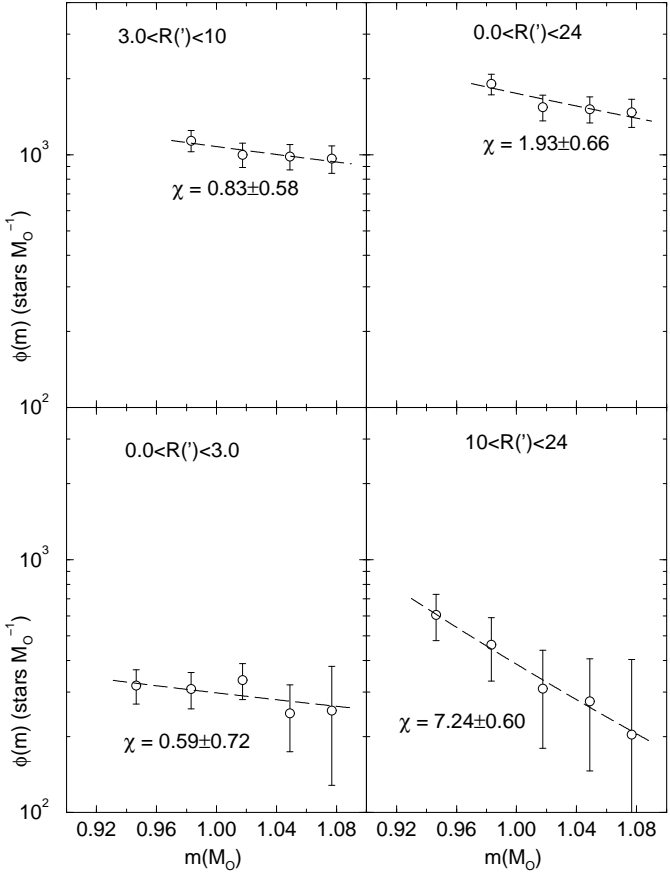}}
\caption[]{Mass functions $\phi(m)$ restricted to the turnoff--turnover region.
The MF fits $\phi(m)\propto m^{-(1+\chi)}$ are shown as dashed lines.}
\label{fig6}
\end{figure}

Cluster mass estimates can be made after converting the turnoff -- turnover LFs into 
MFs according to $\phi(m)=\phi(\mj)\left|\frac{dm}{d\mj}\right|^{-1}$, using the 
stellar mass-luminosity relation taken from the 7.1\,Gyr Padova isochrone. To avoid
the star excess near the turnoff, we restricted this analysis to stars with mass
$\leq1.08\,\ms$ ($\mj\approx2.7$). To increase the statistical representativeness of our 
results, we applied this method to the core region, $\rm 3.0\arcmin\leq R\leq10\arcmin$, 
$\rm 10\arcmin\leq R\leq24\arcmin$ and to the overall LF. The resulting MFs, including
$1\sigma$ Poisson error bars for each data point, are shown in Fig.~\ref{fig6}.

The spatial variation in the stellar content of NGC\,188 is clearly seen in the
MFs shown in Fig.~\ref{fig6}, being flat in the core region and very steep in the
region $\rm 10\arcmin\leq R\leq24\arcmin$. We quantify this effect by fitting the function 
$\phi(m)\propto m^{-(1+\chi)}$ to the MFs, using a weighted non-linear least-squares
routine. The results are given in Table~\ref{tab1}. The number of stars from the high-mass 
end to just below the turnoff ($m=1.08\,\ms$) is obtained by numerical integration of the 
LFs in the corresponding mass interval (co.~4). Then, multiplying this number by the 
mass at the turnoff gives us an estimate of the mass stored in evolved stars (m$^*_{evol}$), 
which is listed in col.~3 of Table~\ref{tab1}. In col.~6 we provide the MS mass, obtained 
by numerical integration of the MFs from the turnover to $m=1.08\,\ms$. Col.~7 gives the 
total observed mass contained in each sampled region of NGC\,188, while the corresponding number 
of observed stars is listed in col.~8. In col.~9 we provide an estimate of the observed  
mass density in each region. Uncertainties in the values of $\chi$ (col.~5) are derived from the 
least-squares fit, and uncertainties in the mass and number of stars come from error propagation. 

\begin{table*}
\caption[]{Mass-function fit and related parameters.}
\begin{tiny}
\label{tab1}
\renewcommand{\tabcolsep}{1.75mm}
\begin{tabular}{ccccccccccccccc}
\hline\hline
&&&&&&&\multicolumn{3}{c}{Observed}&&\multicolumn{3}{c}{Extrapolated}\\
\cline{8-10}\cline{12-14}\\
\multicolumn{2}{c}{Region}&&m$^*_{evol}$&Fit&$\chi$&\mMs&m&N&$\rho*$&&m&N&$\rho*$\\
\cline{1-2}\\
(\arcmin)&(pc)&&(\ms)&(\ms)& &(\ms)&(\ms)&(stars)&$(\ms\,pc^{-3})$&&(\ms)&(stars)&$(\ms\,pc^{-3})$\\
(1)&(2)&&(3)&(4)&(5)&(6)&(7)&(8)&(9)&&(10)&(11)&(12)\\
\hline
Core    &0.0--1.4&&~35&$0.95\rightarrow 1.08$ & $0.6\pm0.7$ & ~$38\pm3$ &~$73\pm3$&$69\pm5$&6.7&&$~609\pm261$&$~2070\pm1470$&$55.5\pm23.8$\\
3.0--10 &1.4--4.8&&109&$0.98\rightarrow 1.08$ & $0.8\pm0.6$ & $105\pm34$ &$213\pm7$&$200\pm5$&0.5&&$2260\pm968$&$~7700\pm5440$&$5.2\pm2.2$\\
10--24 &4.8--11.6&&~49&$0.95\rightarrow 1.08$ & $7.2\pm0.6$ & ~$47\pm2$ &~$94\pm3$&$91\pm4$&0.02&&$1090\pm475$&$~3760\pm2670$&$0.2\pm0.1$\\
Overall&0.0--11.6&&193&$0.98\rightarrow 1.08$ & $1.9\pm0.7$ & $187\pm12$ &$380\pm12$&$333\pm76$&0.1&&$3760\pm1610$&$12800\pm9040$&$0.6\pm0.3$\\
\hline
\end{tabular}
\end{tiny}
\begin{list}{Table Notes.}
\item Col.~1: spatial region expressed in arcmin; col.~2: same as Col.~1 in pc; col.~3: mass 
of stars later than the turnoff; col.~4: mass range to which the 
MF has been fitted; col.~5: MF slope; col.~6: turnoff -- turnover mass calculated from the MF fit; 
col.~7: observed mass, sum of m$^*_{evol}$ and \mMs; col.~8: number of observed stars;
col.~9: observed mass density; col.~10: extrapolated mass; col.~11: extrapolated number
of stars; col.~12: extrapolated mass density.
\end{list}
\end{table*}

The advanced, mass-segregated dynamical state of NGC\,188 is reflected particularly in the 
flat MF slope in the core region ($\chi=0.6\pm0.7$) compared to the very steep slope 
($\chi=7.2\pm0.6$) in the region $\rm 10\arcmin\leq R\leq24\arcmin$. Indeed, the mass segregation 
is fully characterized by the positive gradient, with respect to the distance to the cluster 
center, presented by the MF slopes (Table~\ref{tab1}).

At this point it may be interesting to compare the dynamical mass of NGC\,188 
(Sect.~\ref{Fund_par}) with an independent determination of the mass still locked up in stars
in this cluster by directly extrapolating the derived MF down to the theoretical 
stellar low-mass end $\rm m_{low}=0.08\,\ms$. Thus, according to the MF in Table~\ref{tab1}, the 
total stellar mass in NGC\,188 is $\mtot\sim(1.8\pm0.7)\times10^4\,\ms$. On the other hand, 
Kroupa, Tout \& Gilmore (\cite{KTG1991}) and Kroupa (\cite{Kroupa2001}) presented evidence
that the MFs of most globular and open clusters flatten below $\sim0.5\,\ms$. As a consequence, 
the total stellar mass in NGC\,188 is probably less than the value derived above, since most 
of the stars are expected to be found in the low-mass range. Accordingly, we derive a more 
conservative total mass value for NGC\,188 assuming the universal IMF of Kroupa (\cite{Kroupa2001}), 
in which $\chi=0.3\pm0.5$ for $0.08\,\ms - 0.50\,\ms$, and $\chi=1.3\pm0.3$ for $0.50\,\ms - 
0.98\,\ms$. For $0.98\,\ms - 1.08\,\ms$ we adopt the value derived in this paper, 
$\chi=1.93\pm0.66$. As expected, the resulting total stellar mass drops to 
$\mtot\sim(3.8\pm1.6)\times10^3\,\ms$. This mass value is in close agreement with the mass 
estimated from NGC\,188's galactocentric distance and tidal radius (Sect.~\ref{Fund_par}). 
Following the procedure described above, we calculate for each region of NGC\,188 the 
extrapolated mass, number of stars and corresponding mass density. These quantities are 
listed in cols.~10, 11 and 12 respectively of Table~\ref{tab1}.

Mass segregation in a star cluster scales with the relaxation time, defined as
$\rm\tr=\frac{N}{8\ln N}\tc$, where $\rm\tc=R/\sigma_v$ is the crossing time, 
N is the number of stars and $\rm\sigma_v$ is the velocity dispersion (Binney \& Tremaine 
\cite{BinTre1987}). For NGC\,188, $\rm R\approx11$\,pc (Sect.~\ref{StructAnal}), and we assume 
a typical $\rm\sigma_v\approx3\,\rm km\,s^{-1}$ (Binney \& Merrifield \cite{Binney1998}). Thus, 
for the observed number of stars, $\tr\sim24$\,Myr, and for the extrapolated number of 
stars, $\tr\sim600$\,Myr. Both estimates of \tr\ are much smaller than the age 
of NGC\,188, which is consistent with the presence of mass segregation. Interestingly, 
the first estimate of \tr\ in NGC\,188 was made by van den Bergh \& Sher 
(\cite{vdBS1960}) who from star counts derived a mass of $\sim900\,\ms$ and a radius 
of $\sim6.5\arcmin$, thus resulting in $\tr\sim64$\,Myr.

With an age of about 7\,Gyr, NGC\,188 must have lost a significant fraction of its 
primordial mass through {\it (i)} internal processes, such as mass loss and disruption 
through the evolution of massive stars in the early phases, and evaporation along the
cluster's evolution, and {\it (ii)} external processes such as tidal shocks in the Galactic 
disk and encounters with giant molecular clouds. During the first few million years of a 
massive cluster's life, as the primordial gas is removed from the system, the gravitational 
well changes, decreasing the escape velocity. As a consequence, the high-velocity stars can 
be ejected from the cluster in a fraction depending on the initial cluster configuration, 
star formation efficiency and velocity distribution of the stars (Adams \cite{Adams2000}, 
de la Fuente Marcos \& de la Fuente Marcos \cite{delaF2}). Subsequently, evaporation sets 
in and starts reducing the cluster mass. Examples of population decline with cluster age, 
against evaporation, can be seen in the numerical models (including stellar mass loss) of 
de la Fuente Marcos \& de la Fuente Marcos (\cite{delaF2}). Disk shocking plays an important 
r\^ole not only in reducing a cluster's mass but also in flattening its spatial shape. 
Indeed, Bergond, Leon, \& Guibert (\cite{Bergond2001}) estimate an upper limit of 5--10\% 
for the mass-loss efficiency during a single disk crossing, and observe strong flattening 
in the shapes of NGC\,2287 and NGC\,2548, which they attribute to disk shocking.

With respect to NGC\,188, the present-day number of stars of $\sim10^4$ (Table~\ref{tab1}) 
is already larger than the primordial number of stars in the most-populous models available 
in de la Fuente Marcos \& de la Fuente Marcos (\cite{delaF2}). Consequently, taking into
account {\it (i)} the sharp population decline with time observed in the populous models of 
de la Fuente Marcos \& de la Fuente Marcos (\cite{delaF2}), {\it (ii)} the other mass-reducing
processes discussed in the preceding paragraph and, {\it (iii)} the old age derived for
NGC\,188 ($\approx7\,$Gyr), it is clear that the primordial population of stars in NGC\,188 
must have significantly exceeded $10^4$ stars. Consequently, its primordial mass must have
been larger than $4\times10^3\,\ms$.

\subsection{Binaries in NGC\,188}
\label{Bin}

The results presented in Sects.~\ref{LumF} and \ref{MassF} are based on the simple 
assumption that all stars in the CMD are single objects. However, stars in multiple 
systems, binaries in particular, usually account for a considerable fraction of the 
stellar content in open clusters (e.g. Montgomery, Marschall \& Janes \cite{Montg1993}). 

As a consequence of the dynamical evolution in open clusters, multiple systems tend to 
concentrate in the central regions, thus changing the initial spatial distribution of 
stars (Takahasi \& Portegies Zwart \cite{TakaP2000}). Observationally, the main effect 
of a significant fraction of unresolved binaries in the central parts of a star cluster 
is that the number of low-mass stars is underestimated with respect to the higher-mass 
stars. Thus, the observed central LF (or MF) turns out to be flatter than that of the 
actual, resolved stellar distribution. In addition, the resulting MF changes will depend 
both on the binary fraction and MS mass range (e.g. Kroupa, Tout \& Gilmore \cite{KTG1991} 
and references therein). 

With respect to NGC\,188, von Hippel \& Sarajedini (\cite{Hippel1998}) estimate a fraction 
of $\approx50\%$ of multiple systems in the central region. 
At this point, the natural question to ask is whether this large fraction of binaries 
in NGC\,188 may account for the observed flat central MF (Fig.~\ref{fig6} and
Table~\ref{tab1}).  

To address this question we adopted the following approach. First we generate a population
of single stars according to a  standard ($\chi=1.35$) Salpeter MF distribution. To 
be consistent with the 2MASS CMD of NGC\,188, we used single stars in the mass range 
0.15--1.10\,\ms. Then, according to a pre-defined fraction of CMD binaries, we randomly 
select stars from the original distribution and build the specified number of binaries.  
Mass and luminosity biases are avoided in this process  (Kroupa \cite{Kroupa2002}; Mazeh 
et al. \cite{Mazeh2003}) and consequently, the probability of a star being selected as a 
binary member depends only on the number frequency of its mass. Each star selected to 
become a binary member is subsequently deleted from the original population. Based on the 
7.1\,Gyr isochrone mass-luminosity relation, the combined luminosity of each pair member 
is converted back to mass. When all binaries are thus formed and their masses stored in 
the new population distribution, we add the remaining (not-paired) original single stars 
to the final population distribution. The final, observed MF is then built by counting 
the combined number of single and paired stars in each mass bin. To reach a high level of 
statistical significance, the initial single-star distribution contains $10^6$ stars. 

\begin{figure}
\resizebox{\hsize}{!}{\includegraphics{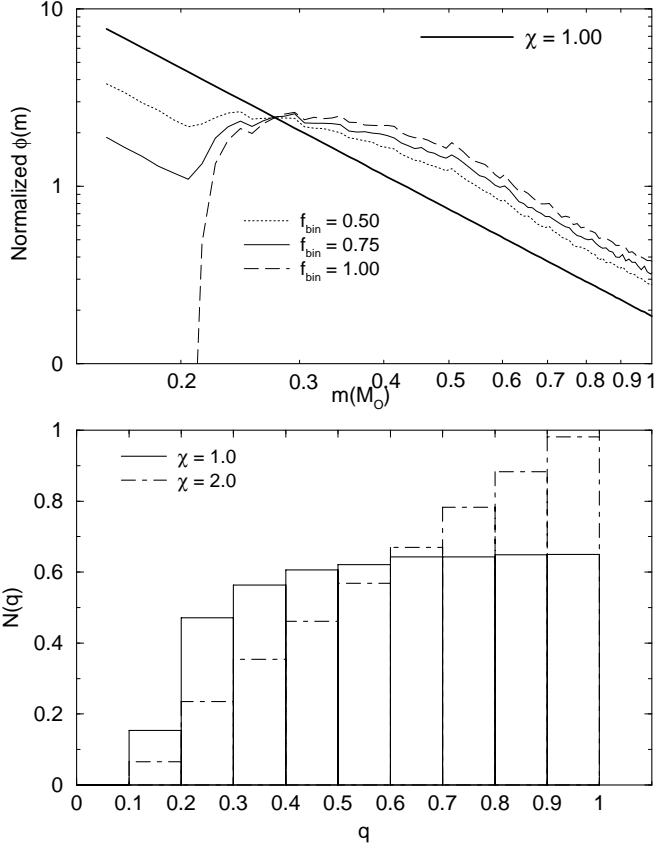}}
\caption[]{Top panel: varying binary fractions are tested  for a single-star MF of 
slope $\cmf=1.0$. The (original) single star population MF is represented by the heavy 
solid line. All MFs are scaled so that they produce the same total number of stars.
Bottom panel: simulated mass-ratio ($q=M_2/M_1$) distribution for MFs with $\cmf=1.0$
and $2.0$. The binary fraction for both distributions is $\fb=0.75$.}
\label{fig7}
\end{figure}

To check how a varying binary population may affect the original, single-star MF, we applied 
the above approach to MFs with slopes $\cmf=1.0, 1.5$ and 2.0, and for fractions of CMD 
binaries $\fb=0.5, 0.75$ and 1.0. $\fb=0.75$ means that 75\% of the observed CMD objects 
are binaries.  We illustrate this effect for $\cmf=1.0$ in the top panel of Fig.~\ref{fig7}, 
in which the heavy solid line represents the MF of the single-star distribution. The final MF 
distributions are shown as dotted, thin solid and dashed lines, respectively for the binary 
fractions $\fb=0.5, 0.75$ and 1.0. All MFs in Fig.~\ref{fig7} are scaled so that the total 
number of stars produced by each MF $\left(\displaystyle\int\phi(m)\,dm\right)$ is the 
same.

Since most of the single-star distribution is made up of low-mass stars, binary formation
severely depletes the original MF, particularly for $m\leq0.25\,\ms$. The binaries thus formed 
have a mass ratio $q\sim0.65$ and tend to accumulate, in varying proportions (according to \fb\ 
and \cmf), in the region $0.3\leq m(\ms)\leq0.9$ of the observed MF. Thus, considering 
the distribution of stars with mass $\geq0.25\,\ms$, the final MFs indeed turn out flatter 
than the single-star distributions. However, for stars in the mass range from which we derived 
the present results (mass $\geq0.8\,\ms$), the final distributions become even steeper than 
the original ones. This is particularly true for the $\cmf=2.0$ MF, since this 
distribution contains the largest fraction of low-mass stars.
   
We summarize the above discussion in Table~\ref{tab2}, in which we give, for each \cmf, 
the resulting slopes obtained from MF fits to the mass ranges $0.2\leq m(\ms)\leq1.1$ and 
$0.6\leq m(\ms)\leq1.1$.

\begin{table}
\caption[]{Binary effects on MFs }
\begin{tiny}
\label{tab2}
\renewcommand{\tabcolsep}{0.40mm}
\begin{tabular}{ccccccccccc}
\hline\hline
&&&\multicolumn{2}{c}{$\fb=0.50$}&&\multicolumn{2}{c}{$\fb=0.75$}&&\multicolumn{2}{c}{$\fb=1.00$}\\
\cline{4-5}\cline{7-8}\cline{10-11}\\
\cmf&q&&$\chi_{0.2-1.1}$&$\chi_{0.6-1.1}$ &&$\chi_{0.2-1.1}$&$\chi_{0.6-1.1}$&&$\chi_{0.2-1.1}$&$\chi_{0.6-1.1}$  \\
\hline
1.0&$0.61\pm0.23$&&0.50&1.12&&0.36&1.14&&0.26&1.15\\
1.5&$0.65\pm0.23$&&0.90&1.79&&0.75&1.88&&0.75&1.92\\
2.0&$0.69\pm0.22$&&1.37&2.33&&1.25&2.43&&1.29&2.50\\
\hline
\end{tabular}
\begin{list}{Table Notes.}
\item \cmf\ is the original MF slope; $\chi_{0.2-1.1}$ is the MF slope in the mass range 
$0.2\leq m(\ms)\leq1.1$ and $\chi_{0.6-1.1}$ is the slope in the range $0.6\leq 
m(\ms)\leq1.1$. 
\end{list}
\end{tiny}
\end{table}

As expected, the $0.2\leq m(\ms)\leq1.1$ MF flattening degree is largest for the flattest 
single-star MF, since the relative number of higher-mass stars with respect to lower-mass 
stars is favoured in this distribution. Indeed, the average binary mass ratio for
$\cmf=1.00$ is, within the uncertainties, the smallest of the MFs tested here. 

Because of the binary formation, the accumulation of stars in the observed MFs for mass 
$\leq0.9\,\ms$ ends up increasing the MF slope in the $0.6\leq m(\ms)\leq1.1$ mass 
range. This effect is present for any \cmf\ and \fb.

Finally, we present in the bottom panel of Fig.~\ref{fig7} the mass-ratio
($q=M_2/M_1$) distribution obtained for MFs with $\cmf=1.0$ and $2.0$, derived
for stars with masses in the range $0.15-1.1\,\ms$. The binary fraction in both 
distributions shown in Fig.~\ref{fig7} is $\fb=0.75$. Since the mass ratio depends 
only on the masses of the two companions, the distributions for different binary
fractions are essentially the same, but vary with different MF slopes. Interestingly, 
the nearly-flat distribution for $\cmf=1.0$ for $0.3<q<1.0$ agrees with the results 
presented in Mazeh et al. (\cite{Mazeh2003}) for spectroscopic binaries  with primary 
masses in the range $0.6-0.85\,\ms$. The similarity between the simulated mass-ratio 
distribution for $\cmf=1.0$ (Fig.~\ref{fig7}) and that observed in spectroscopic binaries
(Mazeh et al. \cite{Mazeh2003}) suggests that Galactic field stars tend to have
an MF with a slope around $\chi=1$.

\subsubsection{Widening of CMD sequences}
\label{photom}

In the magnitude range we are 
dealing with in this work, uncertainties in J and H are typically $\leq0.10$\,mag (e.g.
Soares \& Bica \cite{SB2002}). Despite the small values, the uncertainties produce an 
additional scatter on the observed CMDs and, as a consequence, a widening, particularly 
of the MS. NGC\,188 indeed presents a wide MS (Figs.~\ref{fig1} and \ref{fig4}), and in
this section we investigate how much of the MS width can be attributed to binaries and 
photometric errors. The approach here is to produce synthetic CMDs which simulate
both effects and to compare them with actual CMDs of NGC188.

\begin{figure}
\resizebox{\hsize}{!}{\includegraphics{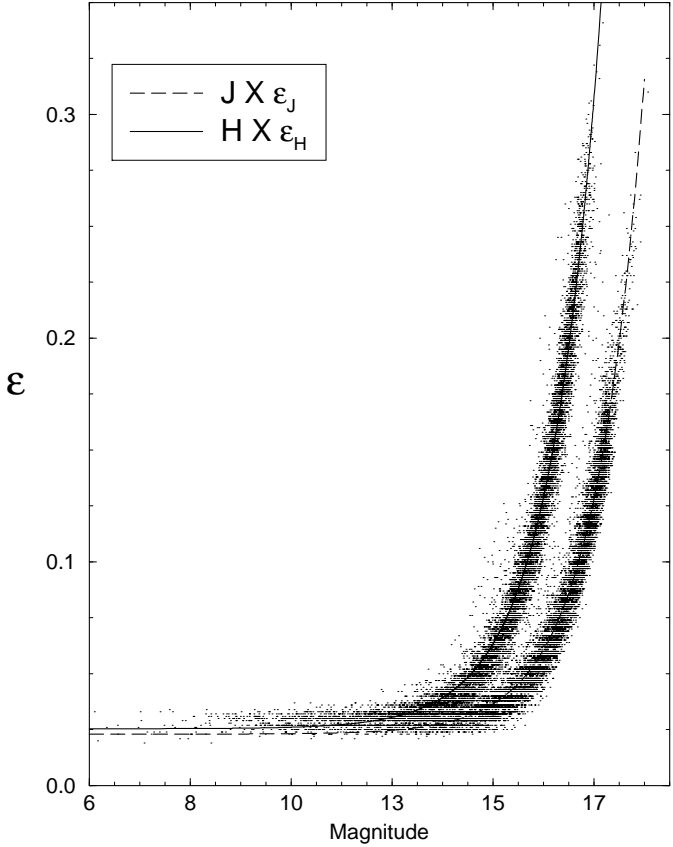}}
\caption[]{Average photometric uncertainty related to a given magnitude. Solid line: $\epsilon_H
\times H$; Dashed line: $\epsilon_J\times\jj$. Least-squares fits result 
in $\epsilon_H=0.025+1.92\times10^{-8}e^{(H/0.998)}$ and 
$\epsilon_J=0.023+4.78\times10^{-9}e^{(J/0.976)}$.}
\label{fig8}
\end{figure}

In 2MASS photometry, the average photometric uncertainty ($\epsilon$) for a given magnitude 
follows a tight relation with magnitude, as can be seen in Fig.~\ref{fig8}, in which we plot
data for NGC\,188. For each J magnitude, the photometric error is obtained by randomly selecting 
a value in the interval ($-\epsilon_J\leq\epsilon\leq+\epsilon_J$), assuming that $\epsilon$ 
follows a normal distribution curve. H errors are obtained similarly.

We first generate an MS stellar distribution according to a given initial mass function 
(IMF) and binary fraction, as described in Sect.~\ref{Bin}. For NGC\,188, the MS is
assumed to follow the 7.1\,Gyr isochrone. For each single-mass star and binary thus formed, 
uncertainties are assigned to the J and H magnitudes as described above. Finally, 
we add to the simulated CMD a background stellar distribution.

\begin{figure}
\resizebox{\hsize}{!}{\includegraphics{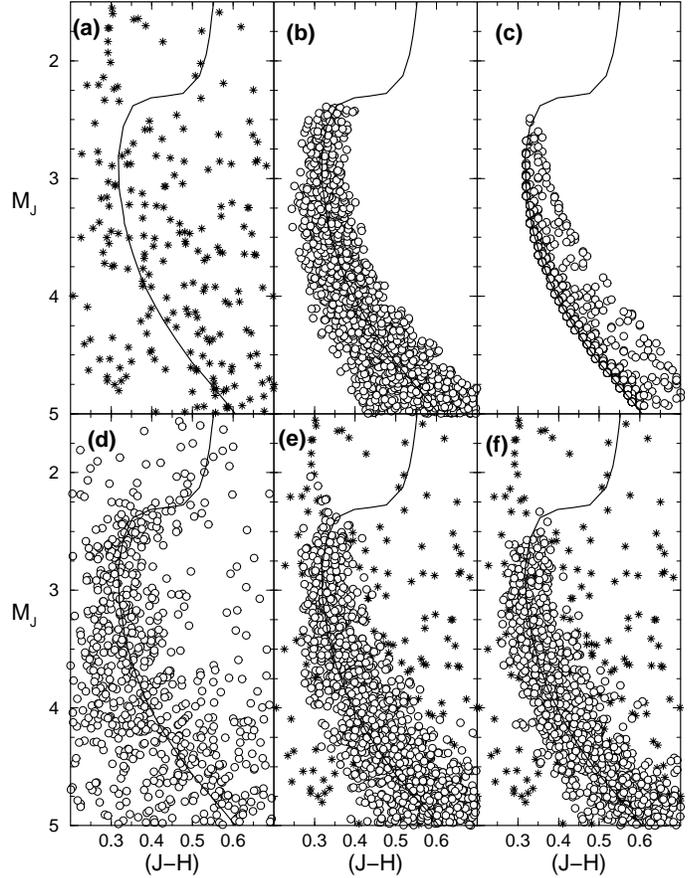}}
\caption[]{Simulated CMDs compared to the observed one. Panel (a) - offset field; (b) - scatter 
produced by 2MASS photometric errors only on a $\chi=0.8$ IMF distribution; (c) - scatter
produced by a binary fraction $\fb=0.5$ on the same IMF of panel (b); (d) -  observed 
CMD in the region R\,$\leq10\arcmin$; (e) -  CMD resulting from the combination of 
photometric errors, binaries ($\fb=1.0$) and the offset field; (f) - same as (e) but for 
$\fb=0.5$. The offset field distribution is shown by the stars.}
\label{fig9}
\end{figure}

We test this approach by comparing the resulting CMDs with the observed one for $\rm R\leq10\arcmin$, 
shown in panel (d) of Fig.~\ref{fig9}; the corresponding (same area) background field is
shown in panel (a). To be consistent with the results of Sects.~\ref{MassF} and \ref{Bin}, we 
use an IMF with $\chi=0.8$ and binary fraction $\fb=0.50$. In addition, the simulated MSs have 
been generated with approximately the same number of stars as the observed, background-subtracted 
MS. We illustrate the individual scatter produced in the IMF distribution by the photometric errors 
and binaries, respectively in panels (b) and (c). The MSs in panels (b) and (c) have similar widths, 
although, as expected, the photometric errors produce scatter symmetrically distributed about the 
isochrone. The binaries, as expected, produce scatter biased towards brighter magnitudes. This 
effect was originally pointed out by Kuiper (\cite{Kuiper1935}). Finally, assigning photometric 
errors to the binaries and single stars of panel (c), and adding the offset field (panel (a)), we 
obtain the comparison CMDs in panels (e) and (f). For illustrative purposes, the 7.1\,Gyr isochrone 
is included in all panels.

The present approach produced near-IR CMD sequence widenings similar to those obtained
in the optical by, e.g. Kerber et al. (\cite{kerber2002}), Hernandez, Valls-Gabaud \& 
Gilmore (\cite{HVGG1999}) and Hurley \& Tout (\cite{HT1998}), in which binaries and 
photometric errors are also included in CMD statistical simulations.

The scatter resulting from the combination of binaries ($\fb=0.50$) and photometric 
errors (panel (f)) produces a thick MS which, in general terms, presents similarities 
with the observed MS (panel (d)). However, the observed MS is clearly wider than the
simulated one. Since the photometric errors are indeed very small in the MS magnitude
range of NGC\,188, the observed excess width is probably due to a larger binary fraction. 
Indeed, a better representation of the observed MS is obtained with $\fb=1.0$ (panel (e)).
Interestingly, Kroupa, Tout \& Gilmore (\cite{KTG1991}) found that for stars to about 
100\,pc from the Sun, the LFs determined both from photometric parallax and counting 
individual stars are consistent with each other if all stars are in binary systems.

In conclusion, the above arguments show that the presence of unresolved binaries in 
the CMD, even in large fractions, cannot account for the large MF flattening degree 
detected in the central parts of NGC\,188. On the other hand, the width of the observed 
MS can be mostly accounted for by a binary population with a fraction $\fb\approx1.0$,
provided that photometric errors are properly taken into account. 

\section{Concluding remarks}
\label{Conclu}

In the present work, the physical structure, stellar content and dynamical state of the old 
open cluster NGC\,188 have been analysed. The spatial dependences of the luminosity and mass
functions were studied in detail. The present analyses made use of J and H 2MASS All Sky data 
release photometry.

NGC\,188 is about 7\,Gyr old and has survived past the age at which most open clusters 
have already dissolved, possibly leaving behind a cluster remnant (Carraro \cite{Carraro2002},
Pavani et al. \cite{Pavani2002},\cite{Pavani2003}). One of the reasons for such a long life 
may be the orbit of NGC\,188, which avoids the inner disk regions most of the time, thus 
decreasing the probability of large-scale encounters with giant molecular clouds. The large 
Galactocentric distance of NGC\,188, $\dgc\approx8.9$\,kpc, also accounts for the long-time 
survival of this open cluster, together with its relatively large mass.

NGC\,188 presents a smooth radial distribution of stars (Sect.~\ref{StructAnal}), with small 
$1\sigma$ Poisson error bars because of the large number of member stars. Its projected radial 
density profile has a marked central concentration of stars (Fig.~\ref{fig2}). From a 
two-parameter King model fit to the projected radial density profile we derive a core radius 
$\rc=1.3\pm0.1$\,pc. The tidal radius, $\rt=21\pm4$\,pc, has been derived by fitting a 
three-parameter King model to the profile. The tidal radius is nearly twice as large as the 
visually determined linear limiting radius of $11.6\pm0.8$\,pc. The concentration parameter
of NGC\,188, $c=1.2\pm0.1$, makes this open cluster structurally comparable to the loose 
globular clusters, but still more concentrated than $\approx25\%$ of them.  The observed 
projected radial density profile of NGC\,188 departs from the two-parameter King model in two 
inner regions, which reflects the non-virialized dynamical state (despite the cluster's old
age) and possibly, some degree of non-sphericity in the spatial shape of this old open cluster. 
Its $\mj\times\jh$ CMD (Fig.~\ref{fig4}), with a well-defined turnoff and giant branch, as well 
as evidence of blue stragglers and binaries, can be fitted with the $7.0\pm1.0$\,Gyr solar 
metallicity Padova isochrones. The best-fit with the 7.1\,Gyr isochrone and related uncertainties 
result in $\mM=11.1\pm0.1$, $\ebv=0.0$ and $\ds=1.66\pm0.08$\,kpc. 

Mass segregation in NGC\,188 is reflected in the spatial variation of its MF (Fig.~\ref{fig7}),
which is flat ($\chi=0.6\pm0.7$) in the core region and steep ($\chi=7.2\pm0.6$) in the 
outskirts. The MF fit $\phi(m)\propto m^{-(1+\chi)}$ resulted in an observed stellar mass 
(MS and giants) of $\approx380\pm12\,\ms$. If one extrapolates the 2MASS MF fit down to the theoretical 
low-mass end $\rm m_{low}=0.08\,\ms$ the total stellar mass in NGC\,188 turns out to be 
$\sim(1.8\pm0.7)\times10^4\,\ms$. However, if we assume a more representative IMF, which flattens 
for masses below $\sim0.5\,\ms$, the total stellar mass in NGC\,188 turns out to be 
$\sim(3.8\pm1.6)\times10^3\,\ms$. This value is in close agreement with the mass
estimated for a tidally truncated cluster at the present position and with the same tidal radius
as NGC\,188.  The primordial mass in NGC\,188 must have been significantly larger
than $\sim4\times10^3\,\ms$, since mass-loss processes such as evaporation and tidal
stripping have been acting on this cluster for about 7\,Gyr.

We also addressed the observational consequences of the presence of unresolved binaries 
in the central parts of NGC\,188, since these systems may dominate the 2MASS CMD. After
testing the effects of different binary fractions and single-star distributions, we conclude
that unresolved binaries alone cannot account for the flatness of the central MF in NGC\,188, 
thus supporting the mass-segregation scenario. Finally, the large width exhibited by the MS of 
NGC\,188 cannot be explained only by photometric errors; instead, it is consistent with a binary 
fraction of nearly 100\%. 

\begin{acknowledgements}
We thank the referee for helpful suggestions.
This publication makes use of data products from the Two Micron All Sky Survey, which is a joint 
project of the University of Massachusetts and the Infrared Processing and Analysis Center/California 
Institute of Technology, funded by the National Aeronautics and Space Administration and the National 
Science Foundation. We employed catalogues from CDS/Simbad (Strasbourg) and Digitized Sky Survey 
images from the Space Telescope Science Institute (U.S. Government grant NAG W-2166) obtained using 
the extraction tool from CADC (Canada). We also made use of the WEBDA open cluster database. We 
acknowledge support from the Brazilian Institution CNPq.
\end{acknowledgements}

%

\appendix

\section{Effects of colour filters and background selection}
\label{cmf}

Since star clusters are usually projected against relatively rich Galactic fields, the observed
CMDs may be heavily contaminated by background stars, which thus introduce additional 
noise in the photometry. This effect is particularly critical for objects projected against 
the Galactic disk or bulge. In this context, a careful subtraction of this contamination, 
usually by means of colour and magnitude filters, is necessary. In this section we will 
examine the consequences of having applied filters to the CMD of NGC\,188, both in its 
physical structure and LF.

\begin{figure}
\resizebox{\hsize}{!}{\includegraphics{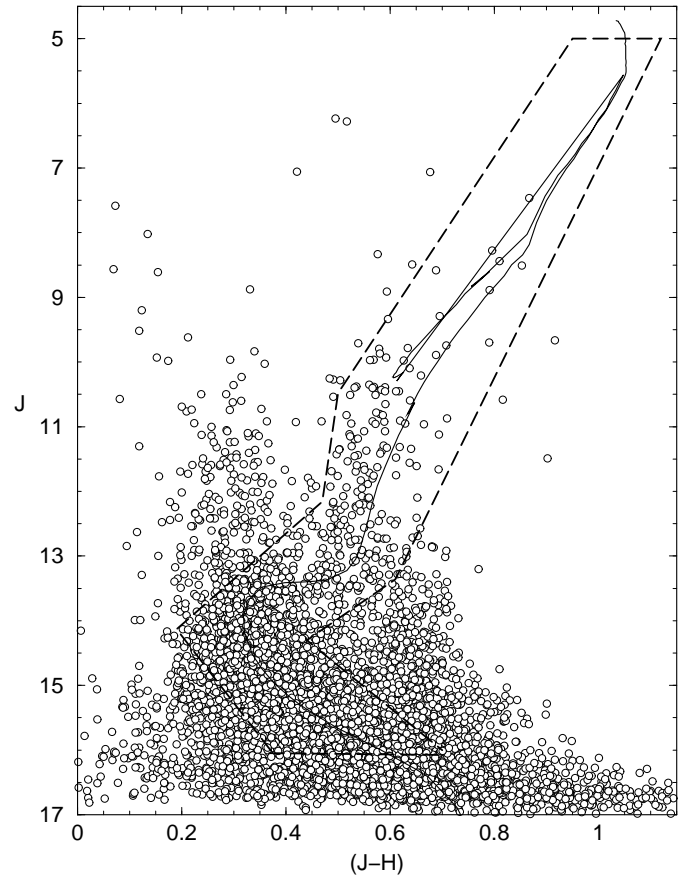}}
\caption[]{The dashed line shows the colour filter used for NGC\,188. Magnitude
cutoffs are $\jj=5$ and $\jj=16$.}
\label{fig10}
\end{figure}

NGC\,188 lies at $b=+22.39^\circ$, consequently its CMD is not heavily contaminated, except 
for disk stars (nearly vertical sequence at $\jh\approx0.2 - 0.4$) and the faint ones
for $\jj\ge16$. Thus, we used a colour filter which follows the 7.1\,Gyr isochrone shape
and preserves most of the stars. The filter is shown as a dashed line in Fig.~\ref{fig10}.

\begin{figure}
\resizebox{\hsize}{!}{\includegraphics{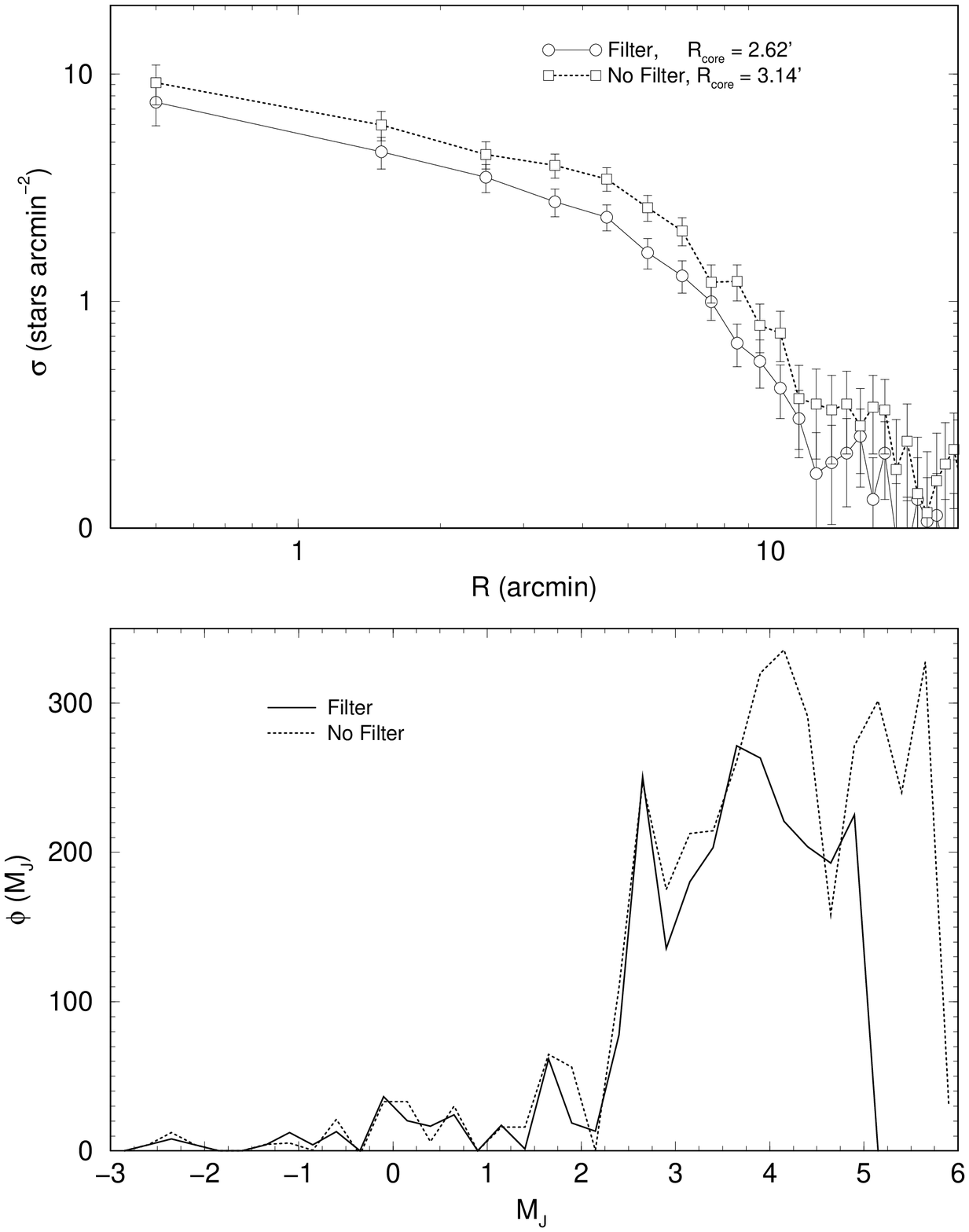}}
\caption[]{Top panel: background-subtracted radial distribution of stars. Bottom 
panel: overall LFs.}
\label{fig11}
\end{figure}

In Fig.~\ref{fig11} (top panel) we compare the radial distribution of stars obtained 
with no filter (circles) and that using the colour filter (squares). The background level 
has been subtracted from both curves. The filtering effect, besides decreasing the number 
of stars, produces a thinner and smoother
profile with respect to that without filter. Indeed, $\rc=2.62\arcmin$ for the
filtered profile while $\rc=3.14\arcmin$ for the unfiltered one.

The overall ($0\arcmin - 22\arcmin$) filtered and unfiltered LFs are shown in the 
bottom panel of Fig.~\ref{fig11}. The two curves are very similar from $\jj=8$ to 
$\jj\approx13.5$. As expected, the differences begin to increase for fainter magnitudes, 
in particular for $\jj\ge14.5$, the magnitude range where most of the discarded stars
are found. If these stars were preserved in the analysis, they would produce an abnormally 
steep profile which would artificially increase the number of low-mass stars in the 
cluster.

We conclude that the use of a colour filter on a CMD does not introduce significant bias 
or artifacts either in the radial distribution of stars or in the LF of a star cluster, 
at least for those clusters resembling NGC\,188. Besides, the changes introduced by
filters are expected to occur in the right direction, since they account for the
background contamination.

Background selection and subtraction is another critical step in the analysis of
star clusters. Ideally, the offset field should represent the background stellar
contribution to the cluster field, both in number and spectral type distribution. 
This is crucial in particular for highly dynamically evolved clusters in which
a significant fraction of the low-mass stars are dispersed in the external regions. 
For these clusters the offset field selection should be guided by a compromise between
distance from the cluster border and representativeness of the background stars.

\begin{figure}
\resizebox{\hsize}{!}{\includegraphics{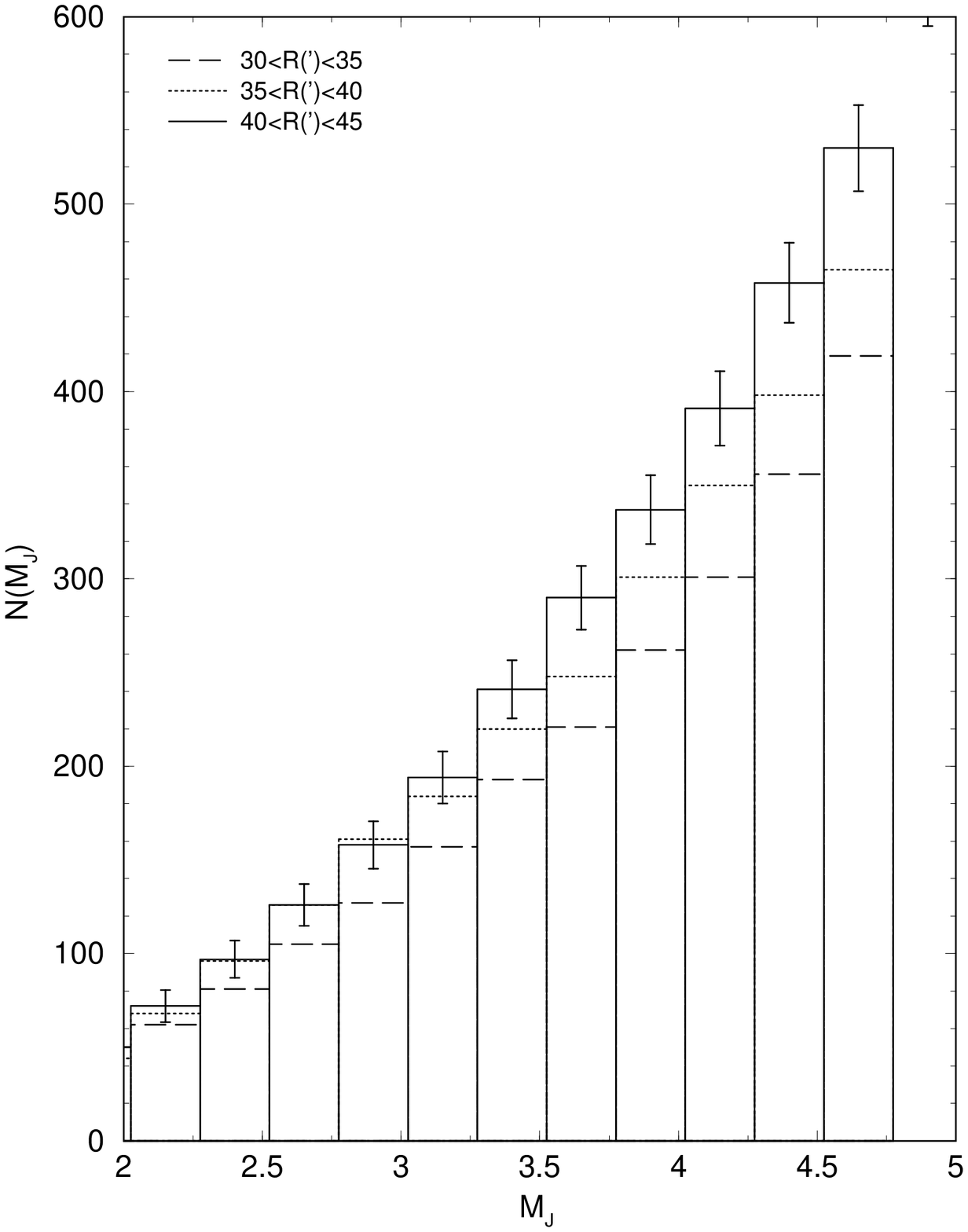}}
\caption[]{Cumulative distribution of stars in external regions of NGC\,188. Dashed 
line: $30\arcmin - 35\arcmin$; Dotted line: $35\arcmin - 40\arcmin$; Solid line: $40\arcmin 
- 45\arcmin$. Poisson error bars for the outermost distribution are shown.}
\label{fig12}
\end{figure}

The uniform sky coverage of 2MASS provides a means to suitably account for the stellar
background. We illustrate this process in Fig.~\ref{fig12}, in which we show the 
cumulative distribution of stars for three circular, external regions of NGC\,188, 
$30\arcmin - 35\arcmin$, $35\arcmin - 40\arcmin$, $40\arcmin - 45\arcmin$. The first 
two distributions have been scaled to match the area of the outermost distribution.

Within the number fluctuation represented by the Poisson error bars, the three distributions 
are similar, which reflects the near uniformity of the stellar content around NGC\,188. 
However, more internal distributions present a slight excess of low-mass stars with respect 
to the outermost distribution, which may be accounted for by the long-term effects of mass 
segregation near the tidal radius. For this reason we selected the region $40\arcmin - 
45\arcmin$ to represent the stellar background for NGC\,188.

\end{document}